\documentclass[prl,twocolumn,showpacs,superscriptaddress,amsfonts,amsmath,floatfix]{revtex4-1}
\usepackage{graphicx}
\usepackage[dvipdfm]{hyperref} 
\usepackage[normalem]{ulem}
\begin{document}
\title{Super Tonks-Girardeau State in an Attractive One-Dimensional Dipolar Gas}
\author{M. D. Girardeau}
\email{girardeau@optics.arizona.edu}
\affiliation{College of Optical Sciences, University of Arizona, Tucson, AZ 85721, USA}
\author{G.E. Astrakharchik}
\email{astrakharchik@mail.ru}
\affiliation{Departament de F\'{\i}sica i Enginyeria Nuclear, Campus Nord B4, Universitat Polit\`ecnica de Catalunya, E-08034 Barcelona, Spain}
\date{\today}

\begin{abstract}
The ground state of a one-dimensional (1D) quantum gas of dipoles oriented perpendicular to the longitudinal axis, with a strong $1/x^3$ repulsive potential, is studied at low 1D densities $n$. Near contact the dependence of the many-body wave function on the separation $x_{j\ell}$ of two particles reduces to a two-body wave function $\Psi_{\text{rel}}(x_{j\ell})$. Immediately after a sudden rotation of the dipoles so that they are parallel to the longitudinal axis, this wave function will still be that of the repulsive potential, but since the potential is now that of the attractive potential, it  will not be stationary. It is shown that as $nd^2\to 0$ the rate of change of this wave function approaches zero. It follows that for small values of $nd^2$, this state is metastable and is an analog of the super Tonks-Girardeau state of bosons with a strong zero-range attraction. The dipolar system is equivalent to a spinor Fermi gas with spin $z$ components $\sigma_{\uparrow}=\perp$ (perpendicular to the longitudinal axis) and $\sigma_{\downarrow}=\parallel$ (parallel to the longitudinal axis). A Fermi-Fermi mapping from spinor to spinless Fermi gas followed by the standard 1960 Fermi-Bose mapping reduces the Fermi system to a Bose gas. Potential experiments realizing the sudden spin rotation with ultracold dipolar gases are discussed, and a few salient properties of these states are accurately evaluated by a Monte Carlo method.
\end{abstract}

\maketitle

Ultracold gases of molecules \cite{Haim,Sage} and atoms \cite{LBL12,AFMBRGF12} with large, permanent magnetic dipole moments have been produced in recent experiments. Such a system is expected to exhibit strong short-range correlations for low densities due to the inverse cube intermolecular potential, and these are strongly enhanced in the 1D limit where transverse excitational energies are large compared with longitudinal translational and thermal energies. In c.g.s. Gaussian units where $\hbar$ and the molecular mass $m$ are $1$, the interaction potential is $d^2/x^3$ if all dipoles are oriented parallel to each other and perpendicular to the longitudinal axis, where $d$ is the dipole moment and $x$ is the molecular separation. For very large dipole moment the system is in the strongly repulsive regime where the relative two-particle wave function vanishes very rapidly as $x\to 0$. We shall show later that the relative wave function for small $x$ is very similar to that of the Tonks-Girardeau (TG) gas \cite{Gir60}, a 1D Bose gas with very strong zero-range repulsion leading to ``fermionization'' of many properties of the system, as recently confirmed experimentally \cite{Par04,Kin04}. Such a configuration is experimentally realizable by applying a sufficiently strong external magnetic field. In the opposite strongly attractive case where all dipoles are oriented parallel to the longitudinal axis, the interaction potential is $-2d^2/x^3$, and we shall show that a highly excited state is produced by sudden rotation of the dipole moments from the transversal to the longitudinal configuration. This state is expected to be metastable, analogous to the ``super Tonks-Girardeau'' (STG) state of a 1D Bose gas with strong zero-range attractions \cite{Ast04,Bat05,Ast05}, which was recently realized in an experiment by the Innsbruck group \cite{Hal09}. Numerical results for the ground state energy of the 1D dipole gas in the repulsive case were found some time ago by the diffusion Monte Carlo method \cite{Ark05},

\emph{Formulation:} Using the above notation, the Hamiltonians of $N$ repulsive or attractive dipoles is
\begin{equation}\label{H}
\hat{H}_\sigma= -\frac{1}{2}\sum_{j=1}^N\frac{\partial^2}{\partial {x_j}^2}+ \sum_{j<\ell}^Nf_\sigma \frac{d^2}{{|x_j-x_\ell|}^3},
\end{equation}
where $\sigma=\perp$ (all dipole moments parallel to each other and perpendicular to the longitudinal axis) or $\parallel$ (all dipole moments in longitudinal direction), $f_\perp =1$, and $f_\parallel =-2$. These dipole moment directions can be treated formally as $z$ components of spin-$\frac{1}{2}$ particles. In the case of fermionic dipoles, this spinor Hamiltonian can be mapped to a spinless Hamiltonian by a Fermi-Fermi mapping which has been used several times previously in different contexts \cite{Gir10,Gir11}. This mapping is
\begin{equation}\label{spinor-scalar map}
\psi_\text{spinor}=M(x_1,\sigma_1;\cdots;x_N,\sigma_N)\psi_\text{spinless}
\end{equation}
where the spin-dependent Fermi-Fermi mapping function $M$, which maps the strongly interacting spinor Fermi gas to a spinless Fermi gas, is
\begin{eqnarray}
&&M(x_1,\sigma_1;\cdots;x_N,\sigma_N)=\prod_{1\le j<\ell\le N}\alpha(x_j,\sigma_j;x_\ell,\sigma_\ell)\nonumber\\
&&\alpha(x_j,\sigma_j;x_\ell,\sigma_\ell)=(\delta_{\sigma_j\perp}\delta_{\sigma_\ell\parallel}
-\delta_{\sigma_j\parallel}\delta_{\sigma_\ell\perp})\text{sgn}(x_j-x_\ell)\nonumber\\
&&\hspace{2.5cm}+\delta_{\sigma_j\perp}\delta_{\sigma_\ell\perp}+\delta_{\sigma_j\parallel}\delta_{\sigma_\ell\parallel}\ ,
\end{eqnarray}
and the signum function $\text{sgn}(x)$ is $+1$ ($-1$) if $x>0$ ($x<0$). This applies to all states $\psi_\text{spinor}$ of the spinor Fermi dipole gas and corresponding mapped state $\psi_\text{F,spinless}$ of the single-component Fermi dipole gas, but we shall restrict ourselves to the ground state. Note that the mapping function $M$ is constructed so as to absorb all of the spin dependence, leaving $\psi_\text{F,spinless}$ spin-free.

\emph{Analytical solutions.} --- The mapped state $\psi_\text{F,spinless}$ represents a spinless Fermi gas. The original 1960 FB mapping is \cite{Gir60} $\psi_\text{B,spinless}=\psi_\text{F,spinless}A(x_1,\cdots, x_N)$ where the ``unit antisymmetric function'' $A$ is $\pm 1$ everywhere and changes sign only at contact points $x_j=x_\ell$ \cite{Gir60}. Numerical results for the ground state energy
(which is exactly the same for fermionic or bosonic repulsive dipoles)
were found previously \cite{Ark05} by the diffusion Monte Carlo method, and a number of other properties of the system were found numerically as well. By a well-known theorem, the Bose ground state is positive everywhere in the $(x_1,\cdots,x_N)$ configuration space except for nodes. This mapping was originally devised to find the exact ground state of the TG gas, for which the interatomic interaction has zero range and is positively infinite at contact. This is very different from the dipole gas, but the $1/(x_j-x_\ell)^3$ divergence of the repulsive dipolar interaction at contact also forces the Bose ground state to zero at contact. It follows that for the ground state the mapping reduces to $\psi_\text{B,spinless}=|\psi_\text{F,spinless}|$ \cite{Note1}. The mapping function $A$ depends only on spin variables except at contact points, where it vanishes, and the Hamiltonian depends only on space variables. It follows that the Hamiltonian commutes with $A$, so the energy is invariant under the mapping. This mapping applies to the case of repulsive dipoles. The lowest state in the perpendicular case is its ground state, whose has density and energy are both zero since the interaction is purely repulsive. If the system is allowed to evolve from a state which is initially localized, it will expand until it reaches density zero
(free expansion), the zero-energy ground state.

Suppose that now, starting with the repulsive ground state, the external magnetic field is rotated suddenly from the perpendicular to the parallel orientation \cite{Note2}. Near contact the dependence of the many-body wave function on the separation $x_{j\ell}$ of two particles reduces to a two-body wave function $\Psi_{\text{rel}}(x_{j\ell})$. Immediately after the sudden rotation, this wave function will still be that of the repulsive potential, but since the potential is now that of the attractive potential, it will not be stationary. It is shown below that as $nd^2\to 0$ where $n$ is the 1D density, the rate of change of this wave function approaches zero. It follows that for small values of $nd^2$, this state is metastable, has energy nearly zero, and is an analog of the super Tonks-Girardeau (STG) state of bosons with a strong zero-range attraction \cite{Note3}.

In the limit where the separation $x_{j\ell}$ of particles $j$ and $\ell$ goes to zero, the dependence of the many-body wave function on $x_j$ and $x_\ell$ reduces to a two-body wave function $\Psi_{\text{rel}}(x_{j\ell})$. In the repulsive case (dipoles parallel to each other and perpendicular to the longitudinal axis), Schr\"{o}dinger's equation for $\Psi_{\text{rel}}$ in c.g.s. Gaussian units is
\begin{equation}\label{repulsive Sch}
\left(-\frac{\partial^2}{{\partial x}^2}+\frac{d^2}{x^3}\right)\Psi_{\text{rel}}(x)=E_{\text{rel}}\Psi_{\text{rel}}(x)\ .
\end{equation}
%
%
%
%
%
In the special case of scattering at zero energy $E_{\text{rel}}=0$ the two-body solution for repulsive (rep.) potential can be written explicitly \cite{NBS}:
\begin{equation}\label{2body solution repulsion}
\Psi_{rep.}(x)=\sqrt{x}K_1(2d/\sqrt{x})
\end{equation}
in terms of a modified Bessel function $K_1$. The short-range expansion $\Psi_{rep.}(x)\propto x^{3/4}\exp({-2d/\sqrt{x}})$ demonstrates that the probability of meeting of two particles is exponentially suppressed due to diverging $d^2/x^3$ interaction. In the limit $d\to 0$ of ideal Fermi/Tonks-Girardeau gas Eq.~(\ref{2body solution repulsion}) can be expanded as $\Psi_{rep.}(x)\propto x$, that is dipolar repulsion suppresses two-body effects much more efficiently than the Pauli principle. The large-distance expansion of Eq.~(\ref{2body solution repulsion}), $\Psi_{rep.}(x) \propto x - d^2\ln[\exp(1-2\gamma)x/d^2]$, has a linear term in $x$ typical for the (unbound) zero-energy scattering state with a logarithmic subleading term. A comparison to a similar expansion for a $\delta$-pseupodotential $\psi(x)\propto x-a_s$ shows that the dipolar gas cannot be described in terms of a $s$-wave scattering length $a_s$.

So far we have assumed that $x>0$. To determine the solution for $x<0$, note that the Hamiltonian is time-reversal invariant since the time-reversal operation reverses both dipole moments and momenta, and the Hamiltonian depends only on $d^2$ and $p^2$, where $p=-i\frac{\partial}{\partial x_j}$. The solution can therefore be chosen to be a parity eigenstate, and we want the even solution since the strong repulsion produces a cusp at $x=0$. Therefore, the solution for all $x$ is obtained by replacing $x$ by $|x|$ in the above.

{\it Attractive case.} --- An analytical expression for a zero-energy eigenstate in the attractive potential can be found by the substitution $d^2\to-2d^2$ in Eq.~(\ref{repulsive Sch}), or, equivalently, $d\to d\sqrt{-2}=\pm id\sqrt{2}$. The two signs for the square root show that this state will be twofold degenerate.
\begin{eqnarray}
\label{2body solution attraction}
\Psi_{attr.}(x)&=&\sqrt{x}K_1(\pm i2\sqrt{2}d/\sqrt{x})
\end{eqnarray}
Taking linear combinations of two solutions it is also possible to express two independent solutions as $\Psi_{attr.}^{(1)}(x)=\sqrt{x}J_1(2\sqrt{2}d/\sqrt{x})$ and $\Psi_{attr.}^{(2)}(x)=\sqrt{x}Y_1(2\sqrt{2}d/\sqrt{x})$. The short-range expansion of Eq.~(\ref{2body solution attraction}) contains an oscillating term $x^{3/4}\exp(\pm i2\sqrt{2}d/\sqrt{x})$ which introduces an infinite number of nodes with an envelope which vanishes as $x^{3/4}$ when $x\to 0$. In other words, the zero-energy solution is a highly excited state of a attractive dipolar gas with an inifinite number of (bound) states below.

The usual sudden-approximation procedure is to expand the state just after the sudden spin rotation (same as the state just before) in terms of energy eigenstates in the new potential. The two-body solutions are physically relevant to the full many-body system up to distances smaller than the interparticle distance $L = n^{-1}$, so the integration in the calculations of the overlap between the original and excited state will be done only over the range $0<|x|<L$.

An important issue is how much the results depend on experimental imperfections. The guiding parameter relevant for the energy is $nd^2$.  It varies linearly with respect to small deviations in density $n$. The interaction strength is changed by a sudden rotation of dipoles from the initial perpendicular orientation with an angle of $\varphi=\pi/2$ to the parallel one $\varphi=0$. The interaction strength depends on the angle $\varphi$ as $(1-3\cos^2\varphi)$ and follows a weak quadratic dependence on small imperfections in the initial and final angles. This situation can be contrasted with that in a usual sTG gas created by crossing the confinement induced resonance, where a small imprecision in the magnetic field can lead to large differences in the interaction strength.

\emph{Metastability.} --- Here it will be shown that the excited gas-like state obtained after a fast rotation of dipoles is metastable for $nd^2\to 0$. A homogeneous system of bosons with attractive interaction is unstable with respect to collapse to a many-body bound state. In fact, there is an infinite number of states lying below the gas-like state for attractive dipoles as discussed above. We also note that the ground state corresponds to a collapsed one as well in the case of Fermi-Dirac statistics. Indeed, the quantum pressure term arising from the Pauli exclusion principle is not enough to avoid the collapse of fermions. In terms of an effective potential, the Pauli principle manifests itself in a repulsive centrifugal $\propto 1/x^2$ potential at short distances, which competes with a stronger attractive $-2/x^3$ dipolar interaction. At the same time it is important to keep in mind that the effective description~(\ref{H}) in terms of a simple dipolar interaction fails when two particles come very close to each other.

The long-range properties in a gas are governed by phonons and hence will not change much in a fast transition between repulsion to attraction as long as the system stays in a gas-like state with positive energy. Differently, the short-range physics is strongly modified. The main effect of the change in the interactions on short-range physics can be analyzed within two-body scattering problem. If there is a large overlap between repulsive (ground) state and attractive gas-like (excited) state, the latter will be efficiently populated during a fast rotation and the system will be metastable.

We analyze the overlap between the two-body solutions for repulsive $d^2/|x|^3$, Eq.~(\ref{2body solution repulsion}), and attractive $-2d^2/|x|^3$, Eqs.~(\ref{2body solution attraction}), interaction potentials. The overlap integral
\begin{eqnarray}\label{overlap integral}
O(L)
=
\sqrt{
\frac{
\int\limits_0^L \Psi_1^*(x)\Psi_0(x)dx
\int\limits_0^L \Psi_0^*(x)\Psi_1(x)dx
}
{
\int\limits_0^L |\Psi_0|^2dx
\int\limits_0^L|\Psi_1|^2dx
}}
\end{eqnarray}
is evaluated up to the mean interparticle distance $L=n^{-1}$.

\begin{figure}
\begin{center}
\includegraphics[width=0.5\columnwidth, angle=-90]{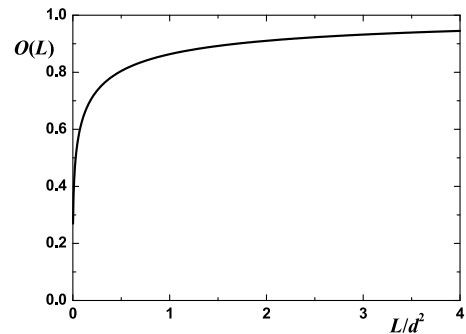}
\caption{Overlap integral (\ref{overlap integral}) between two-body zero-energy scattering solutions for repulsive $d^2/|x|^3$, Eq.~(\ref{2body solution repulsion}), and attractive $-2d^2/|x|^3$, Eq.~(\ref{2body solution attraction}), interactions.
}
\label{Fig:overlap integral}
\end{center}
\end{figure}

Figure~\ref{Fig:overlap integral} shows the overlap integral~(\ref{overlap integral}) calculated up to the distance of $L/d^2$. The overlap integral increases as $L/d^2$ is increased. When the mean interparticle distance is large compared to $d^2$, the overlap integral approaches unity. In other words, the fast rotation will be efficient in creating a metastable state of attractive dipoles when the density is small.
In order to quantify in a many-body simulation which is the region of stability of the gas-like state of attractive dipoles we resort to the  variational Monte Carlo method. We calculate the many-body energy of the system obtained by a fast rotation of dipoles in a sudden approximation.

To do so, we first obtain a good variational description of the ground-state wave function before the rotation, than we use this wave function to get the density dependence of the energy $E$ and the chemical potential $\mu = dE/dN$ after the fast rotation. The obtained state is dynamically unstable if the compressibility is negative, that is when $mc^2 = n \partial \mu/\partial n <0$ \cite{Ast05}.

In the limit of low density or small dipolar moment $nd^2\ll 1$, the ground-state energy per particle of a repulsive dipolar gas approaches TG energy and is quadratic in density $E/N = \pi^2\hbar^2n^2/6m$, while in the opposite regime of a quasi-crystal $nd^2\gg 1$ the energy is cubic in density $E/N = \zeta(3)n^3d^2$. \cite{Ark05}. When dipolar moment is small $nd^2\ll 1$, the potential energy is small compared to the kinetic energy and the fast rotation is expected to produce a stable state with a positive energy. Instead, for large $nd^2$ potential energy dominates and the system with attraction is unstable.

We calculate the energy in a many-body system using variational Monte Carlo method. The variational wave function is chosen in Bijl-Jastrow form $\psi(x_1, \dots, x_N) = \prod\limits_{i<j}f_2(|x_i-x_j|)$. The Jastrow term is chosen to satisfy the two-body scattering problem at short distances $f_2(x) = x^{1/2} I_1(2/\sqrt{x}), x<R_{\text{match}}$ and to be phononic form \cite{Reatto} at large distances $f_2(x) =\sin^\alpha(\pi x/L), x>R_{\text{match}}$. Two asymptotic expressions are matched continuously at the matching distance $R_{\text{match}}$ which fixes the value of $\alpha$ and proportionality constants. The Jastrow term chosen in this way satisfies periodic boundary condition $f_2'(L/2)=0$ with $L = N/n$ being the size of the simulation box of $N$ particles. The variational energy calculated with this function for repulsive interactions is very close to the diffusion Monte Carlo (DMC) energy\cite{Ark05}, which suggests that the quality of the variational function is good. We find it reasonable to use the same variational wave function and average the attractive Hamiltonian over it to simulate fast rotation of the dipoles. The resulting state is metastable (the DMC method would converge to a negative ground state energy) and its actual behavior is expected to depend on the specific way in which the state is prepared. From that we hope to find qualitative agreement with possible future experiments.

\begin{figure}
\begin{center}
\includegraphics[width=0.5\columnwidth, angle=-90]{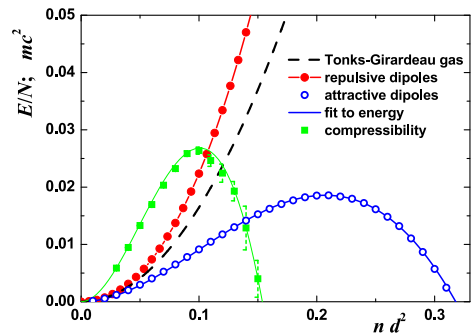}
\caption{(Color online) Energy per particle $E/N$ and compressibility $mc^2$ in units of $\hbar^2/md^2$ as a function of $nd^2$.
Solid red circles, diffusion Monte Carlo energy for repulsive dipoles from Ref.~\cite{Ark05};
open blue circles, variational Monte Carlo energy for attractive dipoles;
solid blue line, polynomial fit $E/N = f(nd^2)\times\hbar^2/md^2$ with $f(x) = a_2x^2+a_3x^3+a_4x^4$ and the coefficients $a_2 = 1.401(6)$, $a_3 = -5.17(5)$, $a_4 = 2.4(1)$ obtained from the best fit to VMC data points;
dashed line, the energy of TG gas $E/N =\pi^2 \hbar^2n^2/6m$;
solid green squares, compressibility $mc^2 = n \partial\mu/\partial n$ in a system of attractive dipoles obtained from numerical derivatives of the VMC energy;
solid green line, polynomial expression $mc^2 = g(nd^2)\times\hbar^2/md^2$ with $g(x) = 6a_2x^2+12a_3x^3+20a_4x^4$ and coefficients $a_2, a_3, a_4$ same as in the fit of the energy.
The energy of a repulsive [attractive] gas is obtained for interaction potential $d/x^3$ [-$2d^2/x^3$] in Hamiltonian~(\ref{H}) for a system of $N=10$ particles.
}
\label{Fig:E}
\end{center}
\end{figure}

The energy dependence on $nd^2$ is shown in Fig.~\ref{Fig:E} for the repulsive and attractive interactions. The energy of a Tonks-Girardeau gas, shown for comparison, is fully kinetic. The potential energy of repulsive dipoles increases further the energy, while attraction lowers it. We find that the energy calculated in sudden approximation becomes negative for $nd^2 \gtrsim 0.3$. By fitting the energy and calculating its derivatives we find that the compressibility becomes negative at $nd^2\approx 0.15$. For larger values of $nd^2$ the system is dynamically unstable. A significantly non-local interaction potential reduces the region of metastability compared to a short-ranged interaction potential. Indeed, in the case of a $\delta$-interaction potential with 1D $s$-wave scattering length $a_{1D}$, Monte Carlo analysis resulted in a positive compressibility for $na_{1D}\lesssim 0.35$ \cite{Ast05} and in an actual experiment the region of stability was even larger \cite{Hal09}.

We note that Bose-Fermi mapping applies also to this metastable state, as it was shown above that the wave function vanishes with an envelope of $x^{3/4}$ when two particles approach each other.

The detection of the sTG state can be best done by observing the frequency of the breathing mode. The signature of sTG state will be an increase in the frequency as the interaction parameters is increased followed by a sudden drop related to the appearance of the instability.

We summarize the stability properties as follows.(i) There is a highly-excited gas-like (sTG) state which is an eigenfunction of the attractive Hamiltonian. In the case of the $\delta$-interaction the wave function of sTG eigenstate was discussed in Ref.~\cite{Gir10-2}. In principle, being an eigenstate, this state has an infinite lifetime. (ii) In practice, the sTG state is created by a fast rotation of dipoles (compare to the fast sweep across the confinement induced resonance in the case of short-range interaction) which populates not only this gas-like state but also lower-energy bound states and which leads to the loss of particles. In the case of short-range interactions, it was demonstrated explicitly in the experimental work of Ref.~\cite{Hal09} that the number of remaining particles decays rapidly for attractive interactions, with almost no decay on the same time scale for repulsive interactions. (iii) We study the metastability of the state obtained after the fast rotation of dipoles by evaluating the energy of the attractive (rotated) Hamiltonian with the wave function of repulsive (non-rotated) Hamiltonian. We find that the energy first increases as the dimensionless parameter $nd^2$ is increased, as it should in a gaseous state, and starts to decrease at larger values of $nd^2$, manifesting the appearance of collapse. The actual values of the energy and the region of the stability in an experiment depend on exactly how the sTG state is populated and quantitatively might be different from our results. (iv) In the region $nd^2>0.15$ we find a negative compressibility, indicating complete instability of the gas-like state, so the sTG state probably cannot be realized in this regime.

{\it Zeeman term.} --- In an actual experiment the sudden rotation of the dipole moments from the perpendicular to the longitudinal configuration is accomplished by sudden rotation of an external magnetic field. This introduces a Zeeman term $-g\mu\mathfrak{H}\hat{S}^z$ in the Hamiltonian, where $\hat{S}^z$ is the total spin in the instantaneous direction of the magnetic field $\mathfrak{H}$. However, there are no spin-spin interactions and the Zeeman Hamiltonian commutes with the original Hamiltonian (\ref{H}). Therefore, the Zeeman term only shifts all state energies by $-Ng\mu\mathfrak{H}f_{\sigma}$ where $N$ is the total number of dipoles and $f_{\sigma}=1$ when all dipoles (therefore, also the external magnetic field) are perpendicular to the longitudinal axis, and $f_{\sigma}=-2$ when all dipoles (therefore, also the external magnetic field) are parallel to the longitudinal axis. The wave functions are not shifted.

G.E.A. acknowledges support from Ministerio de Educacion, Cultura y Deporte (MEC) (Spain) through the Ramon y Cajal fellowship program. The research of M.D.G. was supported by the U.S. Army Research Laboratory and the U.S. Army Research Office under Grant No. W911NF-09-1-0228. We thank Adolfo del Campo for helpful comments.

\end{document}